\documentclass[twocolumn]{revtex4}
\usepackage{graphicx, amssymb}

\begin{document}

\title{The extension of radiative viscosity to superfluid matter}

\thanks{This research is supported by NFSC under Grants No. 11073008.
     $^{**}$Email: ysh@phy.ccnu.edu.cn \\ Telephone number: 13545359902, 13545119781}

\author{PI Chun-Mei$^1$, YANG Shu-Hua$^2$$^{**}$, and ZHENG Xiao-Ping$^2$\\
{\small $^{1}$ Department of Physics and Electronics, Hubei University of Education, Wuhan 430205, P.R.China}\\
{\small $^{2}$ Institute of Astrophysics, Huazhong Normal University,  Wuhan 430079, P.R.China,}  \\
}


\date{Received: , Accepted: }
\begin{abstract}
The radiative viscosity of superfluid $npe$ matter is studied, and
it is found that to the lowest order of $\delta \mu/T$ the ratio of
radiative viscosity to bulk viscosity is the same as that of the
normal matter.\\

\end{abstract}

\maketitle

As one of the most important transport coefficients, bulk
viscosities of simple $npe$ matter, of hyperon matter and  even of
quark matter, both in normal and superfluid states, have been
extensively studied \cite{fin68, saw8901, haensel92, hae00, hae01,
wan84, saw8902, mad92, gup97, lin02, zhe02, zhe04, zhe05, pan06,
sad0701, sad0702,Gus07,Gus08}, for more references see \cite{don07}.

In fact, the mechanical energy of density perturbations is not only
dissipated to heat via bulk viscosity, but also is radiated away via
neutrinos, this was first pointed out by Finiz and Wolf in
1968\cite{fin68}. However, the damping mechanism through neutrinos
is ignored for several decades, until recently, Sa'd and
Schaffner-Bielich \cite{sad09} named this mechanism the radiative
viscosity, and found it is 1.5 times larger than the bulk viscosity
to all Urca processes in the lowest order of $\delta \mu/ T$, both
in nuclear matter and quark matter. Yang et al. \cite{yan09} studied
non-linear effect of radiative viscosity of $npe$ matter in neutron
stars for both direct Urca process and modified Urca process, and
found that non-linear effect will decrease the ratio of radiative
viscosity to bulk viscosity from $1.5$ to 0.5 (for direct Urca
process) and 0.375 (for modified Urca process); which means that for
small oscillations of neutron star the large fraction of oscillation
energy is emitted as neutrinos, but for large enough ones bulk
viscous dissipation dominates.

It's well known that below certain critical temperature nucleons in
neutron star matter will be in superfluid states. The bulk viscosity
of superfluid nucleon matter have been studied \cite{hae00, hae01,
Gus07}, and it turns out that the superfluidity may strongly reduce
bulk viscosity. This paper aims to study the radiative viscosity of
superfluid nucleon matter, in other words, we will give the
relationship between radiative viscosity and bulk viscosity in the
superfluid case.

Both the  bulk viscosity and the radiative viscosity of $npe$ matter
are related to direct Urca process ($n\rightarrow
p+e+\overline{\nu}_{e}$, $p+e\rightarrow n+\nu_{e}$) and modified
Urca process ($n+N\rightarrow p+N+e+\overline{\nu}_{e}$,
$p+N+e\rightarrow n+N+\nu_{e}$) in different conditions. As shown by
\cite{lat91}, the direct Urca process is allowed by the momentum
conservation when $p_{F_{n}}<p_{F_{p}}+p_{F_{e}}$, and for pure
$npe$ matter where $p_{F_{p}} = p_{F_{e}}$, it corresponds to $n_{p}
/ n > 1/9$. This happens if the density is several times larger than
the standard nuclear matter density $\rho_{0}=2.8\times10^{14}g
cm^{-3}$. In the following, we focus on the direct Urca process.

Let us first recall the formulae of radiative viscosity of simple
non-superfluid $npe$ matter \cite{saw8902, sad09}. Considering a
periodic perturbation to the baryon number density
\begin{equation}
n_{b}(t)=n_{b0}+{\rm {R}}e(\delta n_{b} e^{i\omega t}).
\end{equation}
It will lead to a deviation from $\beta$-equilibrium characterized
by
\begin{equation}
\delta \mu=\mu_{p}+\mu_{e}-\mu_{n}=\delta \mu_{p}+\delta
\mu_{e}-\delta \mu_{n},
\end{equation}
where $\mu_{n}$, $\mu_{p}$ and $\mu_{e}$ are the chemical potentials
of the neutrons, protons and electrons. $\delta \mu$ can be
expressed in terms of the variations of two independent variables
$\delta n_{b}$ and $\delta X_{p}$,
\begin{equation}
\delta \mu=C\frac{\delta n_{b}}{n_{b}}+B\delta X_{p},
\end{equation}
where $X_{p}=n_{p}/n_{b}$ is the proton fraction and the coefficient
functions $C$ and $B$ are given by
\begin{eqnarray}
&& C=n_{p}\frac{\partial \mu_{p}}{\partial n_{p}}
+n_{e}\frac{\partial \mu_{e}}{\partial n_{e}}
-n_{n}\frac{\partial \mu_{n}}{\partial n_{n}} ,\\
&& B=n_{b}\left( \frac{\partial \mu_{p}}{\partial n_{p}}
+\frac{\partial \mu_{e}}{\partial n_{e}} +\frac{\partial
\mu_{n}}{\partial n_{n}} \right) .
\end{eqnarray}
To the leading order, the net reaction rate and the increments of
neutrino emissivity due to off-equilibrium could be written as
\begin{equation}
\Gamma_{\nu}-\Gamma_{\overline{\nu}}=-\lambda_{0}
\frac{\xi^{2}}{\delta \mu},
\end{equation}

\begin{equation}
\dot{\cal E}_{\rm loss}={\cal S}_{0}\xi^{2},
\end{equation}
where\cite{yan09,rei95}
\begin{equation}
\lambda_{0} =\frac{714}{457}\epsilon(T,0)\label{gammad},
\end{equation}
\begin{equation}
{\cal S}_{0}=\frac{1071}{457}\epsilon(T,0)\label{epsilond},
\end{equation}
$\xi=\delta \mu/T$, and $\epsilon(T,0)$ is the neutrino emissivity
in equilibrium
\begin{equation}
\epsilon(T,0)=3.3 \times 10^{-14} \left(\frac{x_{p}\rho}{\rho_{0}}
\right)^{1/3}T^{6} {\rm MeV^{5}} .
\end{equation}

For a periodic process, the expansion and contraction of the system
will induce not only the dissipation of oscillation energy to heat,
but also the loss of oscillation energy through an increasing of the
neutrino emissivity. Bulk viscous coefficient $\zeta$ and radiative
viscous coefficient ${\cal R}$ can be defined for the description of
these dissipation mechanisms, respectively \cite{sad09}
\begin {equation}
\langle \dot{\cal E}_{\rm diss}\rangle =-\frac{\zeta}{\tau}
\int_0^{\tau} dt \left(\nabla \cdot \vec v\right)^2,
\label{epsilon-kin}
\end{equation}

\begin {equation}
\langle \dot{\cal E}_{\rm loss}\rangle =\frac{{\cal R}}{\tau}
\int_0^{\tau} dt \left(\nabla \cdot \vec v\right)^2,
\label{epsilon-kin}
\end{equation}
where $\vec v$ is the hydrodynamic velocity associated with the
density oscillations, and $\tau=2\pi/\omega$ is the oscillation
period. Using the continuity equation, one obtains
\begin{equation}
\zeta=-2 \langle \dot{\cal E}_{\rm diss}\rangle \left(
\frac{\upsilon_0}{\Delta \upsilon } \right)^{2} \left(
\frac{\tau}{2\pi}\right)^{2} \label{z},
\end{equation}
\begin{equation}
{\cal R}=2 \langle \dot{\cal E}_{\rm loss} \rangle \left(
\frac{\upsilon_0}{\Delta \upsilon } \right)^{2} \left(
\frac{\tau}{2\pi}\right)^{2} \label{r},
\end{equation}
where the energy dissipation is
\begin{equation}
 \langle \dot{\cal E}_{\rm diss}\rangle=-\int_{0}^{\tau} (\Gamma_{\nu}-\Gamma_{\overline{\nu}})\delta\mu dt,
\end{equation}
and the neutrino emissivity caused by the oscillation is
\begin{equation}
\langle  \dot{\cal E}_{\rm loss}  \rangle=\int_{0}^{\tau} \dot{\cal
E}_{\rm loss} dt .
\end{equation}

Here, we only present the results, for detailed calculations see
\cite{sad09}
\begin{equation}
\zeta = \frac{\lambda C^2}{\omega^2+\left(\lambda B /n_{b}\right)^2}
,
\end{equation}
\begin{equation}
{\cal R} = \left( \frac{{\cal S}_{0}}{\lambda_{0}}\right) \zeta.
\end{equation}
From Eqs. ($\ref{gammad}$)and($\ref{epsilond}$), one gets
\begin{equation}
{\cal R}=\frac{3}{2}\zeta.
\end{equation}
This relation only holds for small oscillations. If the perturbation
amplitude is large enough, the non-linear effect must be taken into
account and this simple relation is no longer correct \cite{yan09}.

Now let us consider the effect of nucleon superfluidity on radiative
viscosity. Assuming

\begin{equation}
\Gamma_{\nu}-\Gamma_{\overline{\nu}}=-\lambda \frac{\xi^{2}}{\delta
\mu},
\end{equation}

\begin{equation}
\dot{\cal E}_{\rm loss}={\cal S} \xi^{2}.
\end{equation}
Apparently,
\begin{equation}
{\cal R} = \left( \frac{{\cal S}}{\lambda}\right) \zeta.
\end{equation}
In the following, we will show how to calculate ${\cal S}/\lambda$
in the lowest order of $\xi$.

In non-beta equilibrium, the net reaction rate of direct Urca
process with nucleon superfluidity is \cite{vil05}

\begin{eqnarray}
 \Gamma_{\nu}-\Gamma_{\overline{\nu}} &=&
\frac{4\pi}{(2\pi)^{8}} T^5\left[ \prod_{j=1}^3 \int
d\Omega_j\right]
\delta(\mathbf{P}_f-\mathbf{P}_i)|M_{fi}|^{2}\prod_{j=1}^3 p_{F_j}
m_j^\ast
\nonumber \\
&&  \int\limits_{0}\limits^{+\infty} d x_\nu
x_\nu^2[J(x_\nu-\xi,v_j)-J(x_\nu+\xi,v_j)], \label{Jpm}
\end{eqnarray}
and the total neutrino emissivity is\cite{yak01,pi09}
\begin{eqnarray}
\epsilon(\xi,v_j)&=& \frac{4\pi}{(2\pi)^{8}} T^6\left[ \prod_{j=1}^3
\int d\Omega_j\right]
\delta(\mathbf{P}_f-\mathbf{P}_i)|M_{fi}|^{2}\prod_{j=1}^3 p_{F_j}
m_j^\ast
\nonumber \\
&&  \int\limits_{0}\limits^{+\infty} d x_\nu  x_\nu^3
[J(x_\nu-\xi,v_j)+ J(x_\nu+\xi,v_j)], \label{Jpm}
\end{eqnarray}
where
\begin{eqnarray}
 J(x,v_j) = \int\limits_{-\infty}\limits^{+\infty} d x_1 d x_2 d x_3 f(z_1)f(z_2)f(x_3)
 \nonumber \\
 \times\delta(z_1+z_2+x_3-x), \label{Jpm}
\end{eqnarray}
and the subscript $j=1,2,3$ corresponds to $n,p,e$ respectively.
$v_j=\frac{\Delta_j}{T}$ is the gap amplitude and
$z_j=\frac{\varepsilon_j-\mu_j}{T}$ (where $j=1,2$),
$x_3=\frac{\varepsilon_e-\mu_e}{T}$.
$x_\nu=\frac{\varepsilon_{\nu}}{T}$ is the dimensionless energy of
the neutrino,  and $f(x)=(1+e^x)^{-1}$ is the Fermi-Dirac functions
of nucleons and electrons, $p_{F_j}$ is the Fermi momentum and
$m_j^\ast$ is the effective particle mass.  $d\Omega_j$ is the solid
angle element in the direction of the particle momentum, and
$|M_{fi}|^2$ is the squared reaction amplitude.

Here, we want to stress that $z_{1}$ and $z_{2}$ in the above two
formula carry all the information about nucleon superfluidity.
Whether for neutron superfluidity or proton superfluidity, near the
Fermi surface we have
\begin{equation} \label{eq:Gap}
      \varepsilon-\mu=\textrm{sign}(\eta)\sqrt{\delta^2 + \eta^2},
\end{equation}
where $\eta=v_{\rm F}(p-p_{\rm F})$, $v_{\rm F}$ and $p_{\rm F}$ are
the Fermi velocity and Fermi momentum, respectively; and
$\delta^2=\Delta^2 F(\vartheta)$, $\Delta$ is the energy gap and
$F(\vartheta)$ describes the dependence of the gap on the angle
$\vartheta$ between the quantization axis and the particle momentum.
For different types of nucleon superfluidity, the expression of
$F(\vartheta)$ is completely different, which can be seen in
\cite{vil05,yak01,pi09}.

In the case of $\xi\ll 1$, to the lowest order we have

\begin{equation}
J(x_\nu-\xi,v_j)-J(x_\nu+\xi,v_j)= -2\xi \partial
J(x_\nu,v_j)/\partial x_{\nu},
\end{equation}

\begin{equation}
J(x_\nu-\xi,v_j)+J(x_\nu+\xi,v_j)- 2J(x_\nu,v_j) = \xi^{2}
\partial^{2} J(x_\nu,v_j)/\partial x_{\nu}^{2},
\end{equation}
then the net reaction rate is
\begin{eqnarray}
\Delta\Gamma&=&\Gamma_{\nu}(\xi,v_j)-\Gamma_{\overline{\nu}}(\xi,v_j)
\nonumber \\
&=& \frac{4\pi}{(2\pi)^{8}} T^5\left[ \prod_{j=1}^3 \int
d\Omega_j\right]
\delta(\mathbf{P}_f-\mathbf{P}_i)|M_{fi}|^{2}\prod_{j=1}^3 p_{F_j}
m_j^\ast
\nonumber \\
&& \times (-4 \xi)  \int\limits_{0}\limits^{+\infty} d x_\nu  x_\nu
J(x_\nu,v_j), \label{gam}
\end{eqnarray}
and the neutrino emissivity due to the departure from
$\beta$-equilibrium is
\begin{eqnarray}
\dot{\cal E}_{\rm loss}&=&\epsilon(\xi,v_j)-\epsilon(0,v_j)
\nonumber \\
&=& \frac{4\pi}{(2\pi)^{8}} T^6\left[ \prod_{j=1}^3 \int
d\Omega_j\right]
\delta(\mathbf{P}_f-\mathbf{P}_i)|M_{fi}|^{2}\prod_{j=1}^3 p_{F_j}
m_j^\ast
\nonumber \\
&& \times 6 \xi^{2} \int\limits_{0}\limits^{+\infty} d x_\nu  x_\nu
J(x_\nu,v_j). \label{enu}
\end{eqnarray}
Note that, the leading order of $\Delta\Gamma$ is the first order of
$\xi$, while for $\dot{\cal E}_{\rm loss}$ it is the second order of
$\xi$.

Unlike normal matter, Eqs.($\ref{gam}$) and ($\ref{enu}$) haven't
exact analytical solutions. However, one can find that $\Delta\Gamma
\delta\mu$ is in proportion to $\dot{\cal E}_{\rm loss}$, thus we
can easily obtain
\begin{equation}
{\cal S}/\lambda=\frac{3}{2},
\end{equation}
which is the same as ${\cal S}_{0}/\lambda_{0}$. This means that the
nucleon superfluidity doesn't change the value of ${\cal R}/ \zeta$.

In summary, we have studied the radiative viscous coefficient of
superfluid $npe$ matter, and find that for direct Urca process, the
ratio of radiative viscosity to bulk viscosity in the lowest order
of $\xi$ yielded by Sa'd and Schaffner-Bielich \cite{sad09} for
normal nucleons could be extended to the superfluid case. In fact,
it can be seen from our calculations that this is correct for
modified Urca process, too. Thus, to the lowest order of $\xi$, the
relation factor of $\frac{3}{2}$ between the radiative viscosity and
the bulk viscosity of $npe$ matter is a generic one.

The following are the extensive discussions to our result:

First, Although we have ${\cal S}/\lambda={\cal S}_{0}/\lambda_{0}$
in the lowest order, the viscosities in the superfluid case are
different from these in the non-superfluid case. The bulk and
radiative viscous coefficients could be easily expressed as
$\zeta=\zeta_{0} R_{\Gamma}$ and ${\cal R}={\cal
R}_{0}R_{\epsilon}$, where $\zeta_{0} $ and ${\cal R}_{0}$ are the
viscosities in the non-superfluid case,
$R_{\Gamma}=\Delta\Gamma(\xi,v_j)/\Delta\Gamma(\xi)$ and
$R_{\epsilon}=\dot{\cal E}_{\rm loss}(\xi,v_j)/\dot{\cal E}_{\rm
loss}(\xi)$ are the reduction factors caused by superfluidity. Of
course, our result shows that to the lowest order of $\xi$,
$R_{\epsilon}$ equals to $R_{\Gamma}$ (note that $R_{\Gamma}$ has
been calculated numerically by Haensel et al. \cite{hae00, hae01}).

Second, our result satisfies the general relationship $\partial
\dot{\cal E}_{\rm loss}/ \partial \delta
\mu=3(\Gamma_{\overline{\nu}}-\Gamma_{\nu})$, which has been found
by Flores-Tuli\'{a}n and Reisenegger \cite{flo06}. They found that
this relationship holds both in the case of normal nucleons and in
the case of superfluid nucleons, and  both in linear case and for
precise solutions.

Finally, our result based on the fact that to the lowest order of
$\xi$, the existence of superfluidity don't change the relationship
between the neutrino emissivity due to the departure from
$\beta$-equilibrium and energy dissipation in the form of heat.
Nevertheless, it is not correct for higher order calculations. As
shown in \cite{yan09,rei95}, for normal matter $\Delta\Gamma$ and
$\dot{\cal E}_{\rm loss}$ can be solved analytically as polynomials
of $\xi$, while for superfluid matter they must be solved
numerically. As a result, we expect that in non-linear regime, the
ratio of radiative viscosity to bulk viscosity of superfluid matter
no longer equals to that of normal matter. However, the calculation
in the non-linear case is far more complicated and we will consider
it in our further study.

\end{document}